\documentclass[conference]{IEEEtran}
\IEEEoverridecommandlockouts
\usepackage{cite}
\usepackage{amsmath,amssymb,amsfonts}
\usepackage{algorithmic}
\usepackage{graphicx}
\usepackage{textcomp}
\usepackage{xcolor}
\usepackage{subfiles}
\usepackage{float}
\usepackage{hyperref}
\usepackage{tikz}

\usepackage{tikz,xcolor,hyperref}

\definecolor{lime}{HTML}{A6CE39}
\DeclareRobustCommand{\orcidicon}{%
	\begin{tikzpicture}
	\draw[lime, fill=lime] (0,0)
	circle [radius=0.16]
	node[white] {{\fontfamily{qag}\selectfont \tiny ID}};
	\draw[white, fill=white] (-0.0625,0.095)
	circle [radius=0.007];
	\end{tikzpicture}
	\hspace{-2mm}
}

\foreach \x in {A, ..., Z}{%
	\expandafter\xdef\csname orcid\x\endcsname{\noexpand\href{https://orcid.org/\csname orcidauthor\x\endcsname}{\noexpand\orcidicon}}
}

\def\BibTeX{{\rm B\kern-.05em{\sc i\kern-.025em b}\kern-.08em
    T\kern-.1667em\lower.7ex\hbox{E}\kern-.125emX}}
\begin{document}

\title{Design Of Two Stage CMOS Operational Amplifier in 180nm Technology
}

\author{\IEEEauthorblockN{1\textsuperscript{st} Tong Yuan}
\IEEEauthorblockA{\textit{School of Microelectronics} \\
\textit{Southern university of Science and Technology}\\
Shenzhen, China \\
yuant2018@mail.sustech.edu.cn\\
}
\and
\IEEEauthorblockN{1\textsuperscript{st} Qingyuan Fan \orcidB{}}
\IEEEauthorblockA{\textit{School of Microelectronics} \\
\textit{Southern university of Science and Technology}\\
Shenzhen, China \\
fanqy2018@mail.sustech.edu.cn\\
}
}

\maketitle

\begin{abstract}
In this paper a CMOS two stage operational amplifier has been presented which operates at 1.8 V power supply at 0.18 micron (i.e., 180 nm) technology and whose input is depended on Bias Current. The op-amp provides a gain of 63dB and a bandwidth of 140 kHz for a load of 1 pF. This op-amp has a Common Mode gain of -25 dB, an output slew rate of 32 $V / \mu s$, and a output voltage swing. The power consumption for the op-amp is $300\mu W$.
\end{abstract}


\section{Introduction}

The trend towards low voltage low power silicon chip systems has been growing due to the increasing demand of smaller size and longer battery life for portable applications in all marketing segments including telecommunications, medical, computers and consumer electronics. The operational amplifier is undoubtedly one of the most useful devices in analog electronic circuitry. Op-amps are built with different levels of complexity to be used to realize functions ranging from a simple dc bias generation to high speed amplifications or filtering. With only a handful of external components, it can perform a wide variety of analog signal processing tasks. Op-amps are among the most widely used electronic devices today, being used in a vast array of consumer, industrial, and scientific devices. Operational Amplifiers, more commonly known as Op-amps, are among the most widely used building blocks in Analog Electronic Circuits.

Op-amps are linear devices which has nearly all the properties required for not only ideal DC amplification but is used extensively for signal conditioning, filtering and for performing mathematical operations such as addition, subtraction, integration, differentiation etc . Generally an Operational Amplifier is a 3-terminal device.It consists mainly of an Inverting input denoted by a negative sign, ("-") and the other a Non-inverting input denoted by a positive sign ("+") in the symbol for op-amp. Both these inputs are very high impedance. The output signal of an Operational Amplifier is the magnified difference between the two input signals or in other words the amplified differential input. Generally the input stage of an Operational Amplifier is often a differential amplifier.

An operational amplifier is a DC-coupled differential input voltage amplifier with an rather high gain. In most general purpose op-amps there is a single ended output. Usually an op-amp produces an output voltage a million times larger than the voltage difference across its two input terminals. For most general applications of an opamp a negative feedback is used to control the large voltage gain. The negative feedback also largely determines the magnitude of its output ("closed- loop") voltage gain in numerous amplifier applications, or the transfer function required. The op-amp acts as a comparator when used without negative feedback, and even in certain applications with positive feedback for regeneration. An ideal Opamp is characterized by a very high input impedance (ideally infinite) and low output impedance at the output terminal(s) (ideally zero).to put it simply the op- amp is one type of differential amplifier. This section briefly discusses the basic concept of op-amp. An amplifier with the general characteristics of very high voltage gain, very high input resistance, and very low output resistance generally is referred to as an op-amp. Most analog applications use an Op-Amp that has some amount of negative feedback. The Negative feedback is used to tell the Op-Amp how much to amplify a signal. And since op-amps are so extensively used to implement a feedback system, the required precision of the closed loop circuit determines the open loop gain of the system.

% 上面这段是复制过来的，到时候要改一下

For this design process, we will first demonstrate the formula of main properties of an operational amplifier in \textbf{Section \ref{section:TA}}, than we will introduce how we find the proper parameters for our design in \textbf{Section \ref{section:DP}}, the simulation result of out design will be presented in \textbf{Section \ref{section:design}}.

\section{Theoretical Analysis}
\label{section:TA}
\subsection{MOSFET ans Two Stage amp}

For MOSFET we have several basic parameters including $$i_D = \frac{1}{2} k_n (\frac{W}{L}) (V_{GS}-V_{TN})^2$$ and $$g_m = \sqrt{2 k_n (\frac{W}{L})} \cdot \sqrt{I_D}$$ for calculation, we have parameters $k_n = 170 \mu A / V^2$ and $k_p = 36 \mu A / V^2$

\subsection{Gain, Pole and zeros}

% 这里只是一个大概，要讲清楚哪个管子对应图片里面的哪一个，然后加上

We define the input $V_{in}$, the output voltage of the first stage i.e. the input voltage of the second stage $V_{1}$, and the output voltage of the whole circuit $V_{out}$, so we can get that for two stage operational amplifier we have $$\frac{V_{out}}{V_{n}} = \frac{V_{out}}{V_{1}} \times \frac{V_{1}}{V_{in}}$$ so we can calculate the voltage gain of two stage separately and then combine together.

We set the output resistance of the first stage $R_{o2} \parallel R_{o4}$ as $R_1$ and the output resistance of the second stage $R_{o6} \parallel R_{o7}$ as $R_2$. We also se the output capacitance of the first stage as $C_1$ and $C_2 \approx C_L$ for the second stage. So we finally get that $$\frac{V_{out}}{V_{in}} = \frac{g_{m1} R_1 \times g_{m2} R_2 \times (1 - \frac{sC_c}{g_{m2}})}{as^2 + bs + 1}$$ with $$a = R_1 R_2(C_1 C_2 + C_1 C_L + C_2 C_L)$$ $$b = R_2(C_c + C_2) + R_1(C_c + C_1) + C_c g_{m2} R_1 R_2$$ and $$g_{m1} = \sqrt{2K_p (\frac{W}{L})_1 I_{D1}}$$ $$g_{m2} = \sqrt{2K_n (\frac{W}{L})_6 I_{D6}}$$ to find the poles and zeros, we must transform the equation into form like $$\frac{V_{out}}{V_{n}} = \frac{A_{dc}(1-\frac{s}{z_1})}{(1+\frac{s}{p_1})(1+\frac{s}{p_2})}$$
here for this two stage amplifier we have the DC gain of the amplifier $$A_{dc} = g_{m1} R_1 \times g_{m2} R_2$$ the zero point of the circuit $$z_1 = \frac{g_{m2}}{C_c}$$, and with a external resistor, $$z_1 = \frac{1}{C_c(\frac{1}{g_{m2}}-R_z)}$$ so we could set $$R_z = \frac{1}{g_{m2}}$$ When it comes to the poles of the circiut, approximately we have $$p_1 \approx \frac{1}{b}$$ we can simply it to $$ p_1 \approx \frac{1}{C_c g_{m2} R_1 R_2}$$ for another pole $p_2$ we have $$p_2 \approx \frac{g_{m2} C_c}{C_1 C_2 + C_1 C_L + C_2 C_L} $$ $C_1$ is very small so we can simply it into $$p_2 \approx \frac{g_{m2}}{C_1 + C_2}$$

% 再画张图（电路图+小信号分析图，带s的交流小信号）

\subsection{Phase Margin}

The gain band with GBW is equal to $DC_{gain} \times p_1 = \frac{g_{m1}}{C_c}$

For phase margin, we have $$\angle \frac{V_{out}}{V_{in}} = -\arctan(\frac{\omega}{z}) -\arctan(\frac{\omega}{p_1})-\arctan(\frac{\omega}{p_2})$$ and we have $$z = 10 \times GBW$$ by substituting $$\angle \frac{V_{out}}{V_{in}} = -\arctan(\frac{GBW}{z}) -\arctan(\frac{GBW}{p_1})-\arctan(\frac{GBW}{p_2})$$  so $$\angle \frac{V_{out}}{V_{in}} = -\arctan(\frac{1}{10}) -\arctan(A_{DC})-\arctan(\frac{GBW}{p_2})$$ then we need $$p_2 \geqslant 2.2 GBW $$ and finally $$C_c > 0.22 C_L$$ to get more than $60^\circ$ phase margin. Thus we also have $$\frac{g_{m1}}{g_{m2}} \leqslant 0.22$$

\subsection{Slew Rate}
In our design, the slew rate is just equal to $$slew rate = \frac{I_5}{C_c}$$ we already have $I_5 = 100\mu A$ so $C_c$ must be under 10C, with is certain to full-fill. Here we need to obtain $10MV/s$ slew rate under 100MHZ, so we need the voltage change more than 0.05V in one pulse, which is 5ns in width.

% 这里的计算没有加补偿电阻，实际上应该不太一样

\subsection{Power}

The power of the op-amp can be calculated by $$I_{total} \times V_{dd}$$

\section{Design Procedure}
\label{section:DP}
\subsection{Design Goal}

\begin{table}[htbp]
\caption{The Design Goal of the operational amplifier}
\begin{center}
\begin{tabular}{|c|c|}

\hline
\textbf{Parameters} & \textbf{Design Goal} \\
\hline
Process & 0.18µm CMOS \\
\hline
$V_{DD}$ & 1.8V \\
\hline
$V_{SS}$  & 0V \\
\hline
Load &  1pF  \\
\hline
Phase margin &  $\geqslant 60^\circ$ \\
\hline
$A_{DM0}$  &  $\geqslant 1000 V/V $ (60dB) \\
% (low-frequency open-loop small-signal gain)
\hline
$A_{CM0}$ &  $\leqslant 0.1 V/V$ (-20dB) \\
% (low-frequency open-loop small-signal common-mode gain)
\hline
Unity gain frequency  & $\geqslant 100MHz$ \\
\hline
Slew rate  & $\geqslant 10V/\mu s$ \\
% \hline
% Nominal input and output common-mode voltage &  NA \\
\hline
Output voltage swing (differential peak to peak) &  $\geqslant 800mVpp$ \\
\hline
Power  &  Minimum \\
\hline
\end{tabular}
\label{tab1}
\end{center}
\end{table}

\textbf{Bonus:} Design your opamp such that the specifications are met under a ±10\% variation of the supply voltage.

\subsection{Design Principle}

The minimum size of the MOSFET we can use is 180nm in length and 400 nm in width, but normally we don't use the minimum channel length due to the increase of the $\lambda$. $L \geqslant 2L_{min} $ is recommended, in this design, we use $L = 1u$. And after initially designed, to optimise the performance of the op-amp, we will adjust the length of some MOSFET while keep the (W/L) unchanged.

To control the systematic offset we set $$\frac{(W/L)_3}{(W/L)_6} = \frac{(W/L)_4}{(W/L)_6} = \frac{(W/L)_5}{2\times(W/L)_7}$$

We also have $$\frac{(W/L)_8}{(W/L)_5} = \frac{I_{ref}}{I_5}$$ and $$\frac{(W/L)_8}{(W/L)_7} = \frac{I_{ref}}{I_7}$$

During the procedure of the design. we first calculate the proper value of the compensate capacitance and resistance, then we will design the first stage, finally the second stage will be designed.

% 后期不行的话就改1微米

\subsection{Parameter Optimization}

\subsubsection{\textbf{Design of $\mathbf{C_c}$}}

To satisfy the phase margin of $60^\circ$ we need $C_c \geqslant 0.22 C_L$, since we have $C_L = 1 \mathbf{ pF}$ so we can use $C_c\geqslant 220 \mathbf{ fF}$. To achieve slew rate $10V/|mu s$ we need $C_c = 10pf$ , to meet a balance between two requirement, and we choose $C_c = 3pf$

% 800这个数据基本就定了，如果其它部分要妥协的话可以按照需要缩小一点
% 这里的计算没有加补偿电阻,但是看王的报告好像不用加

\subsubsection{\textbf{Design of M1 and M2}}

We have $$g_{m1} = GBW \times C_c \times 2\pi$$ and GBW is also called unity gain frequency, which is listed in the design goal with value of 100MHZ. So we need to apply that $$g_{m1} = 100M HZ \times 5pF \times 2\pi = 302 \mu$$ and for convince we choose a litter larger value $510\mu$. Since $$\frac{W}{L} = \frac{g_m ^2}{\mu_n C_{ox} \times 2I_D}$$ and $2I_D = I_5 = 100 \mu A$, fianly we get $$(\frac{W}{L}) = 14.8$$ so we use 20 as the final value of the ratio.

\subsubsection{\textbf{Design of M3 and M4}}

To get more than 800mV of the output range, we need to at least 800mV input common mode voltage range before the zero point, where the gain is 1. This characteristic parameter can also be used to determined the size of the MOSFET M1 and M2. We have $$(\frac{W}{L})_{1,2} = \frac{2I_{D3}}{\mu_p C_{OX} [V_{DD} - ICMR(+) - V_{TH1} + V_{TH3}]^2}$$ we choose ICMR(+) at 1.6V and we get $(\frac{W}{L})_{3,4} \approx 50$.

\subsubsection{\textbf{Design of M5 and M8}}

In the mean while, we also need to fit the proper value of ICMR(-) to determine the size of MOSFET M5. We have $$(\frac{W}{L})_5 = \frac{2U_{D5}}{\mu C_ox (V_{Dsat})^2}$$ with $$V_{Dsat} = ICMR(-) - \sqrt{\frac{2I_{D1}}{\beta_1}} - V_{TH1}$$ Approximately we can choose $(\frac{W}{L})_5 = 30$ and $(\frac{W}{L})_8 = 10$

\subsubsection{\textbf{Design of M6}}

And also we need $g_{m2} \geqslant \frac{g_{m1}}{0.22} $, so we need $g_{m2} \geqslant 2318 \mu$, we want $$V_{DS3} = V_{DS4} = V_{DS6}$$ and $$V_{GS3} = V_{GS4} = V_{GS6}$$  So we need $$\frac{(\frac{W}{L})_6}{(\frac{W}{L})_4} = \frac{I_6}{I_4} = \frac{g_{m2}}{g_{m4}}$$ so here we get $$I_6 = 2.5 \times I_4$$ $(\frac{W}{L})_6 = 150$

% 有点离谱

\subsubsection{\textbf{Design of M7}}

$$\frac{(\frac{W}{L})_7}{(\frac{W}{L})_5} = \frac{I_7}{I_5} = \frac{g_{m7}}{g_{m5}}$$

$$(\frac{W}{L})_7 = (\frac{W}{L})_5 = 10$$

\subsubsection{\textbf{Design of $\mathbf{R_z}$}}

$$R_z = \frac{1}{g_{m2}} = 5k$$

%共模与差模放大倍数
\subsubsection{\textbf{Common and differential mode gain}}

After initially determining the parameters of the MOSFETs, we need to check output voltage gain, than we may need to adjust the parameters to meet the requirements of common and differential mode voltage gain.

\section{Our design}
\label{section:design}
After we initially determine the parameters, we use Parameter Analysis in Cadence Virtuoso to optimize our design and finall we get the design as shown in Table \ref{design}, and the simulation results are in Table \ref{result}.

The final design schematic is named \textit{project-final}

\begin{table}[htbp]
\caption{The Parameters of MOSFETs}
\begin{center}
\begin{tabular}{|c|c|c|c|}

\hline
\textbf{Device} & \textbf{Length (L)} & \textbf{Width (W)} & \textbf{W/L} \\
\hline
M1 & 1u & 20u & 20\\
 \hline
M2 & 1u & 20u & 20 \\
 \hline
M3 & 1u & 50u & 50\\
 \hline
M4 & 1u & 50u & 50\\
  \hline
M5 & 1u & 39u & 39\\
  \hline
M6 & 180n & 20u & 111\\
  \hline
M7 & 1u & 30u & 30\\
  \hline
M8 & 1u & 10u & 10\\
  \hline

\end{tabular}

\begin{tabular}{|c|c|}

\hline
\textbf{Parameter} & \textbf{Design} \\
\hline
$C_c$ & 2pF \\
 \hline
$R_c$ & $7K \Omega$ \\
 \hline
I & 10uA\\
 \hline
\end{tabular}

\label{design}
\end{center}
\end{table}

% 仿真步骤：
% 波特图 放大倍数，相位
% 阶跃函数仿 slew rate
% 大的摆动测输出范围

\begin{table}[htbp]
\caption{The Design result}
\begin{center}
\begin{tabular}{|c|c|c|c|}

\hline
\textbf{Parameter} & \textbf{Target} & \textbf{Achieved} & \textbf{Plot} \\
\hline
Phase margin &  $\geqslant 60^\circ$ & $61.8^\circ$ & Fig \ref{phasemargin}\\
\hline
$A_{DM0}$  &  $\geqslant 1000 V/V $ (60dB) & 67.5dB & Fig \ref{Adm}\\
% (low-frequency open-loop small-signal gain)
\hline
$A_{CM0}$ &  $\leqslant 0.1 V/V$ (-20dB) & -20.9dB & Fig \ref{Acm}\\
% (low-frequency open-loop small-signal common-mode gain)
\hline
Unity gain frequency  & $\geqslant 100MHz$ & 131.9MHZ & Fig \ref{phasemargin}\\
\hline
Slew rate (Rise) & $\geqslant 10V/\mu s$ & $29.7V/\mu s$ & Fig \ref{slewrateraise}\\
\hline
Slew rate (Fall) & $\geqslant 10V/\mu s$ & $12.6V/\mu s$ & Fig \ref{slewratefall}\\
\hline
Output voltage swing &  $\geqslant 800mVpp$ & 936mV & Fig \ref{slewrateraise}\\
\hline
Power & Minimum & $204 \mu W$ & Fig \ref{DC}\\
\hline

\end{tabular}

\label{result}
\end{center}
\end{table}

\section{Conclusion}

In this design, we have satisfied all the parameters in the requirement and specially we achieved high $A_{dm}$, slew rate and wide unity gain phase margin. By comparison we found that the simulation result is a little different from out theoretical design due to some omitting during our calculation. But after all, our calculation has represent the real situation and offered great help in the design of the device.

\section{Appendix}

% \begin{figure*}[htbp]
% \centerline{\includegraphics[width=\textwidth]{design.png}}
% \caption{Design of two stage op amp.}
% \label{design}
% \end{figure*}

\begin{figure}[H]
\centerline{\includegraphics[width=0.5\textwidth]{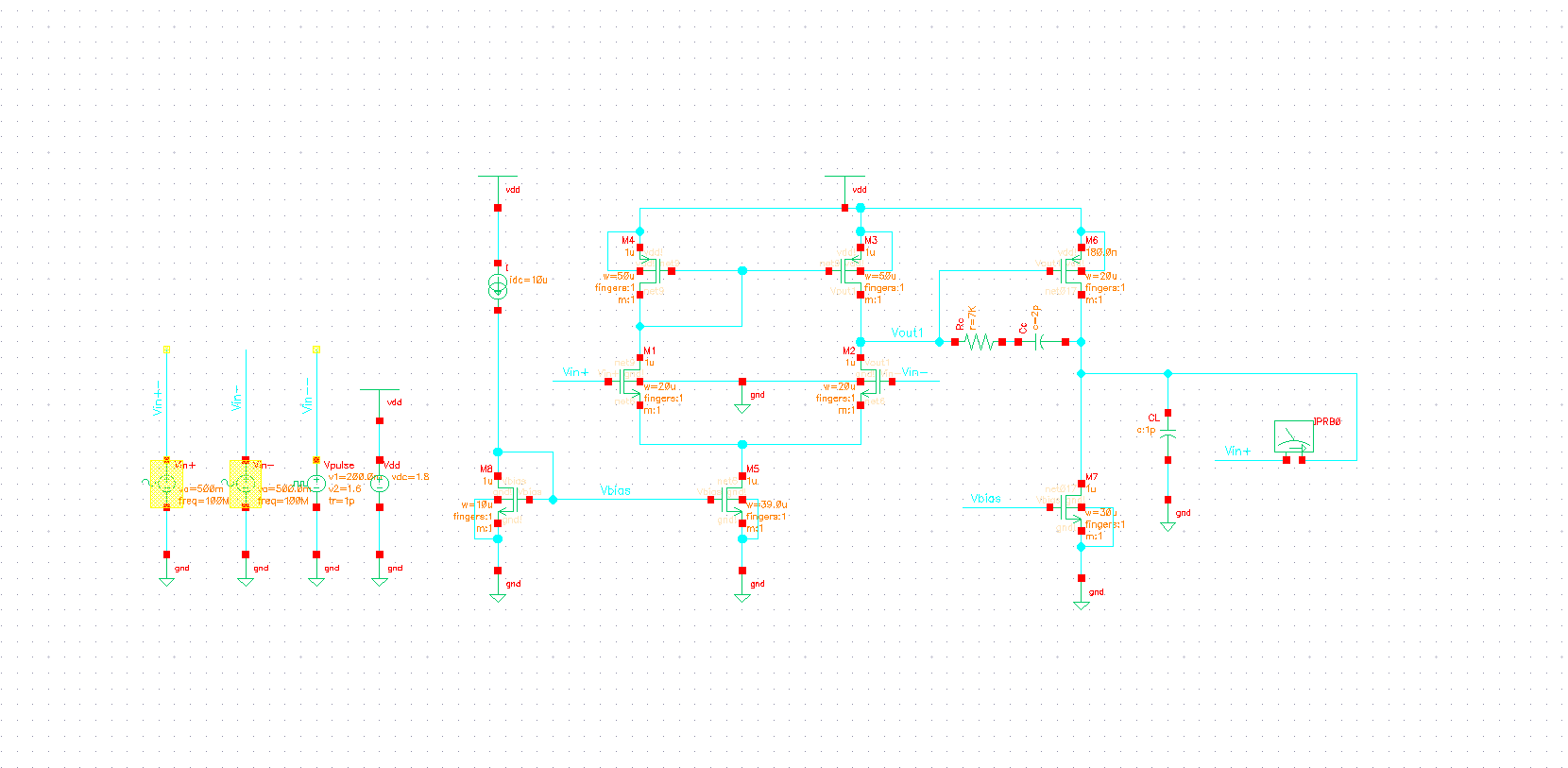}}
\caption{Design of two stage op amp.}
\label{fig}
\end{figure}

\begin{figure}[H]
\centerline{\includegraphics[width=0.5\textwidth]{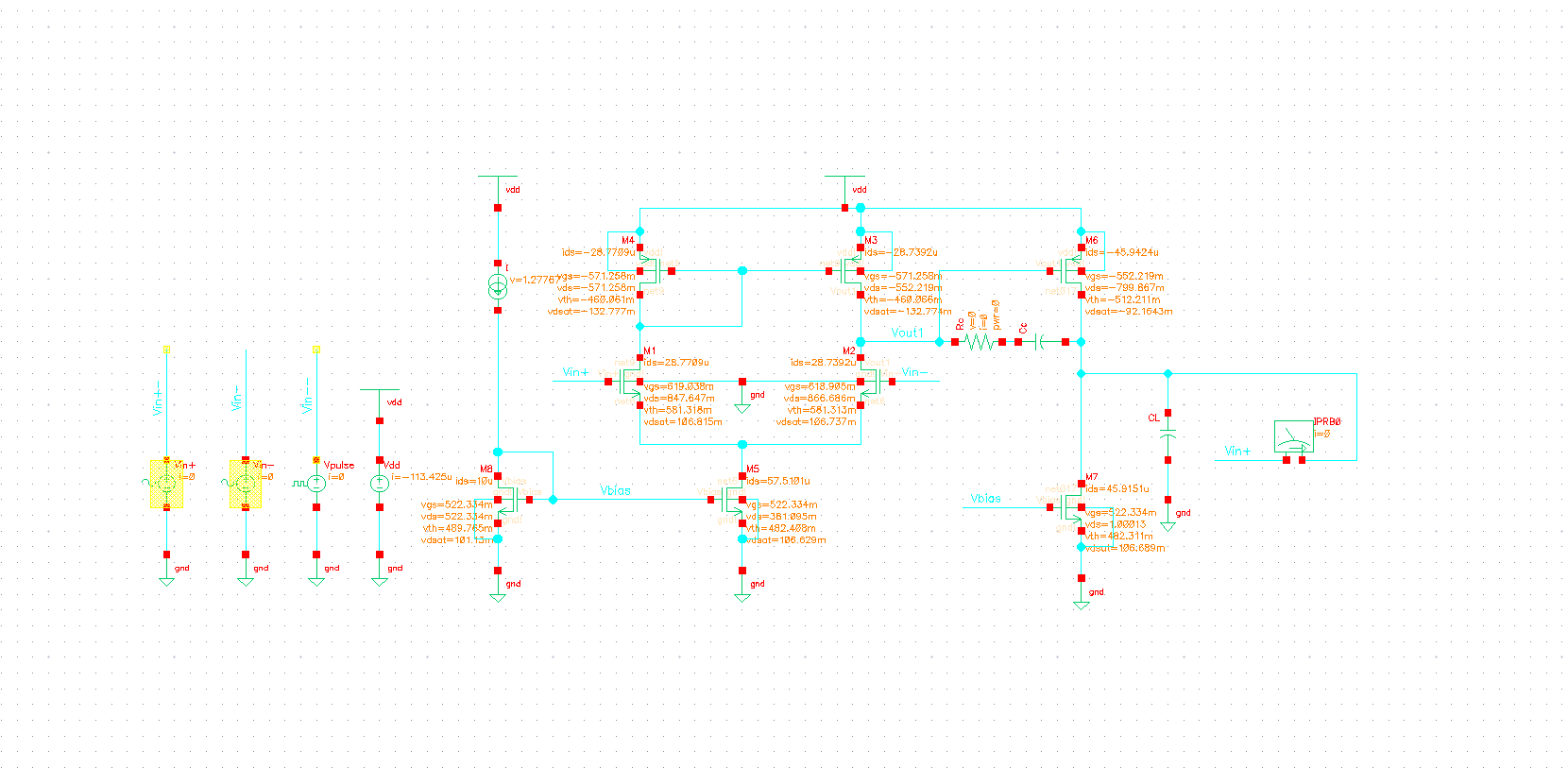}}
\caption{The DC operation point of the amplifer}
\label{DC}
\end{figure}

\begin{figure}[H]
\centerline{\includegraphics[width=0.5\textwidth]{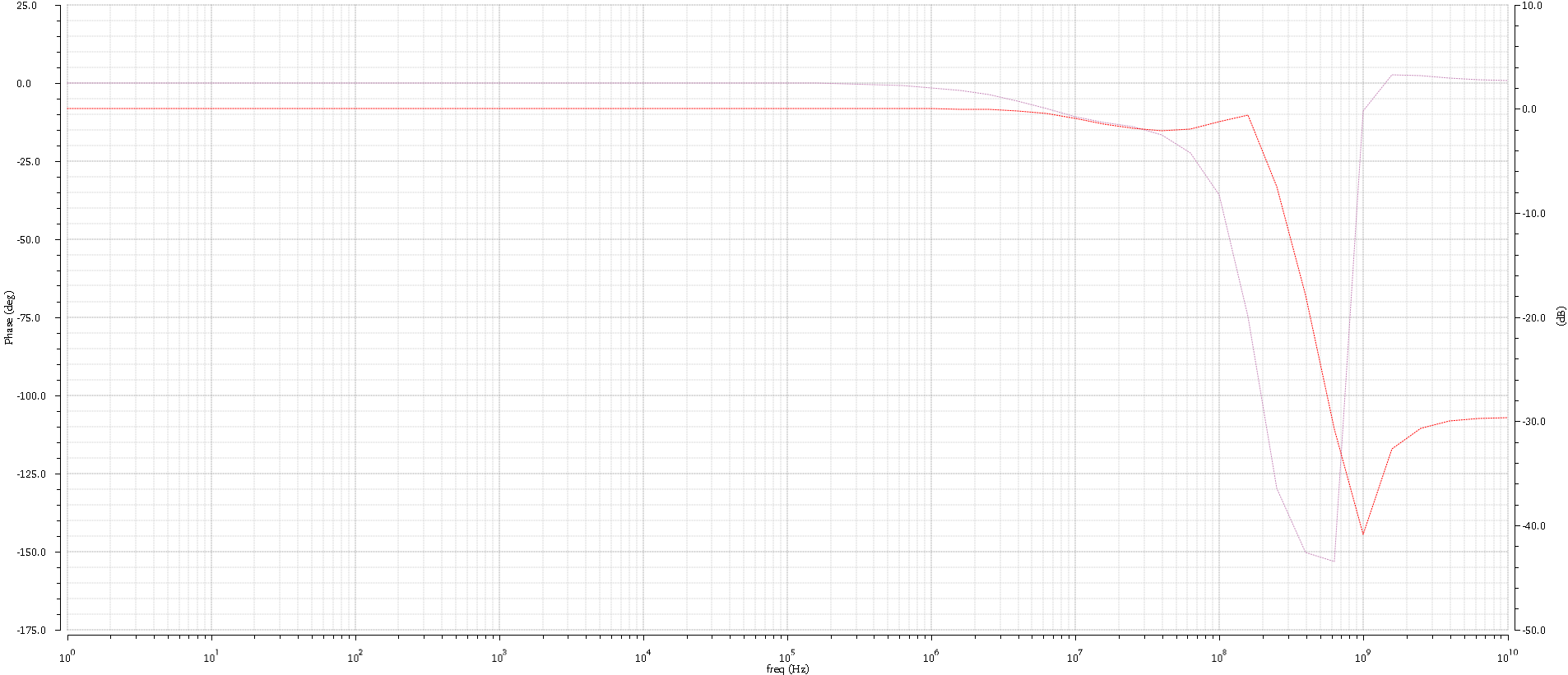}}
\caption{The diagram of the -3dB}
\label{3dB}
\end{figure}

\begin{figure}[H]
\centerline{\includegraphics[width=0.5\textwidth]{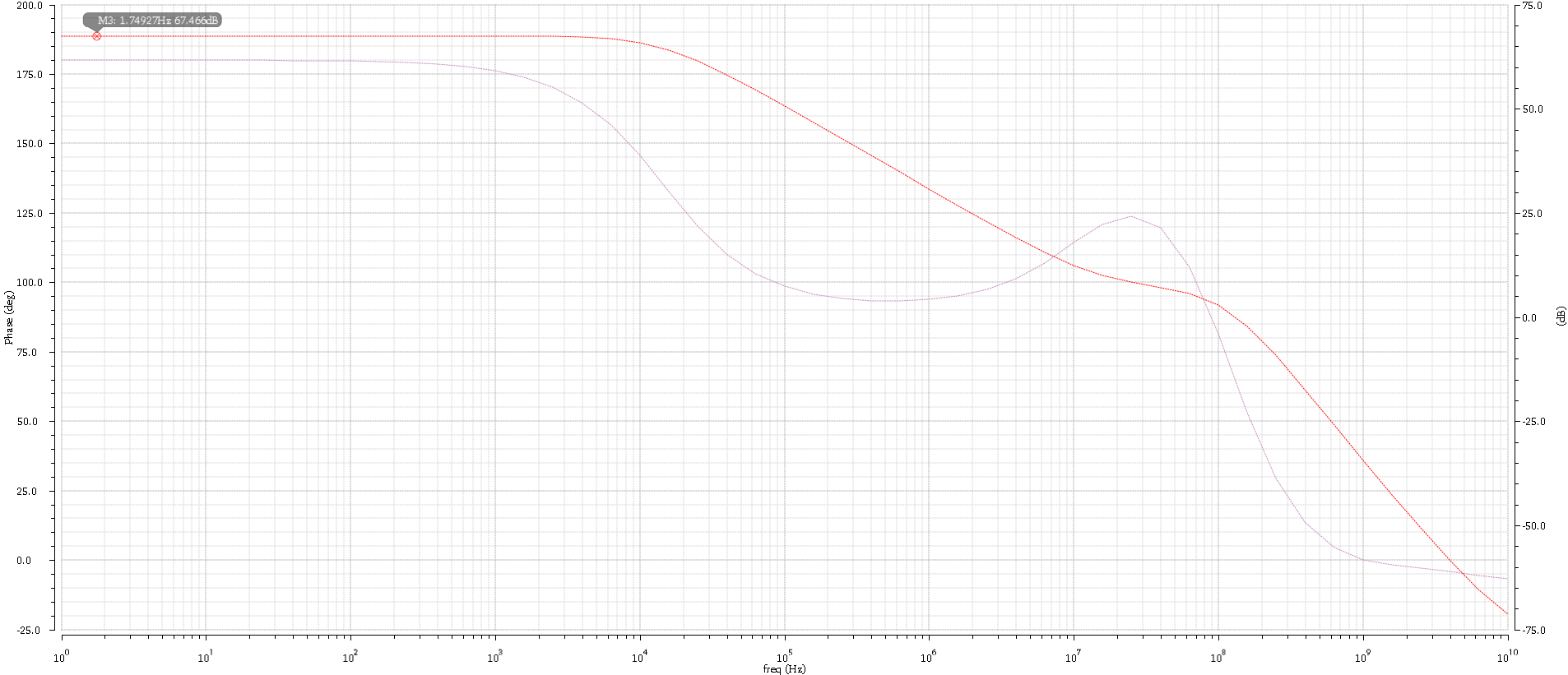}}
\caption{The open loop differential mode gain}
\label{Adm}
\end{figure}

\begin{figure}[H]
\centerline{\includegraphics[width=0.5\textwidth]{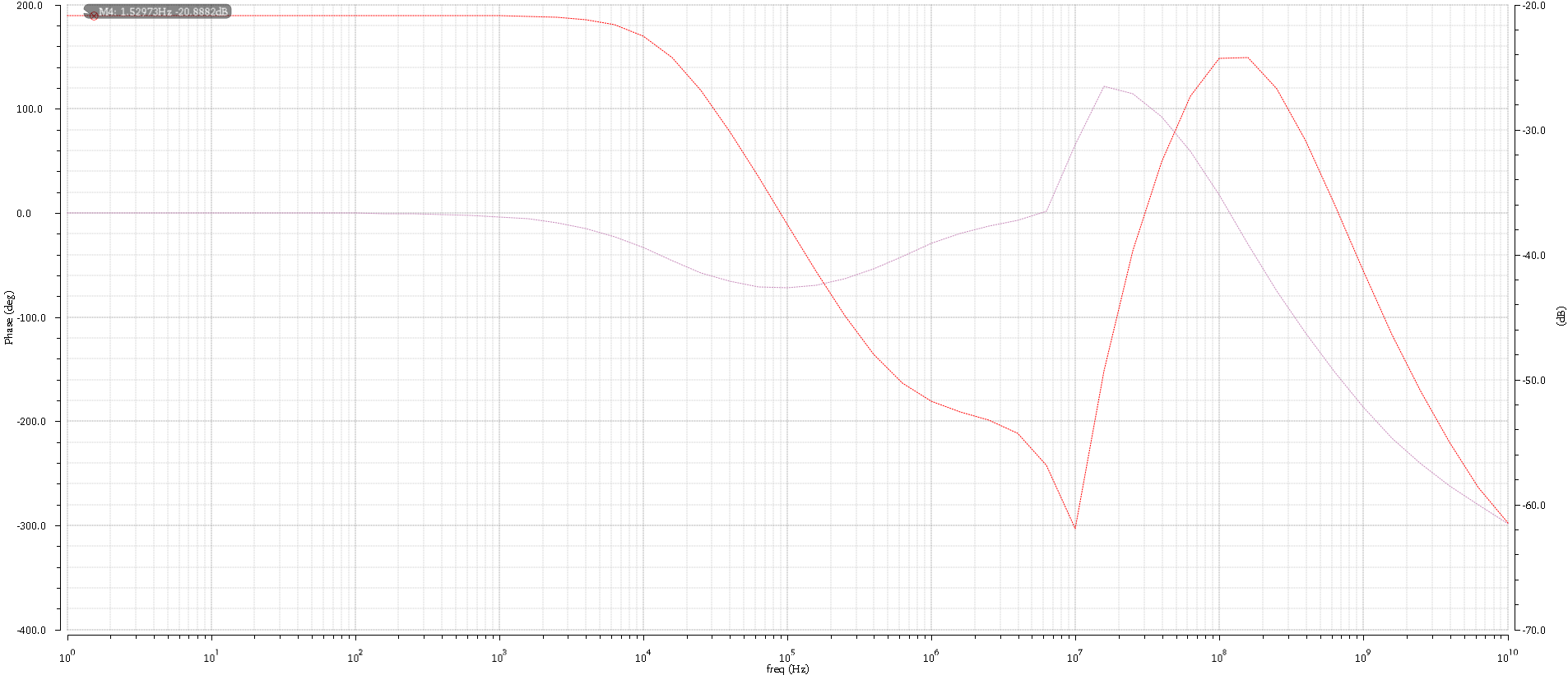}}
\caption{The open loop common mode gain}
\label{Acm}
\end{figure}

\begin{figure}[H]
\centerline{\includegraphics[width=0.5\textwidth]{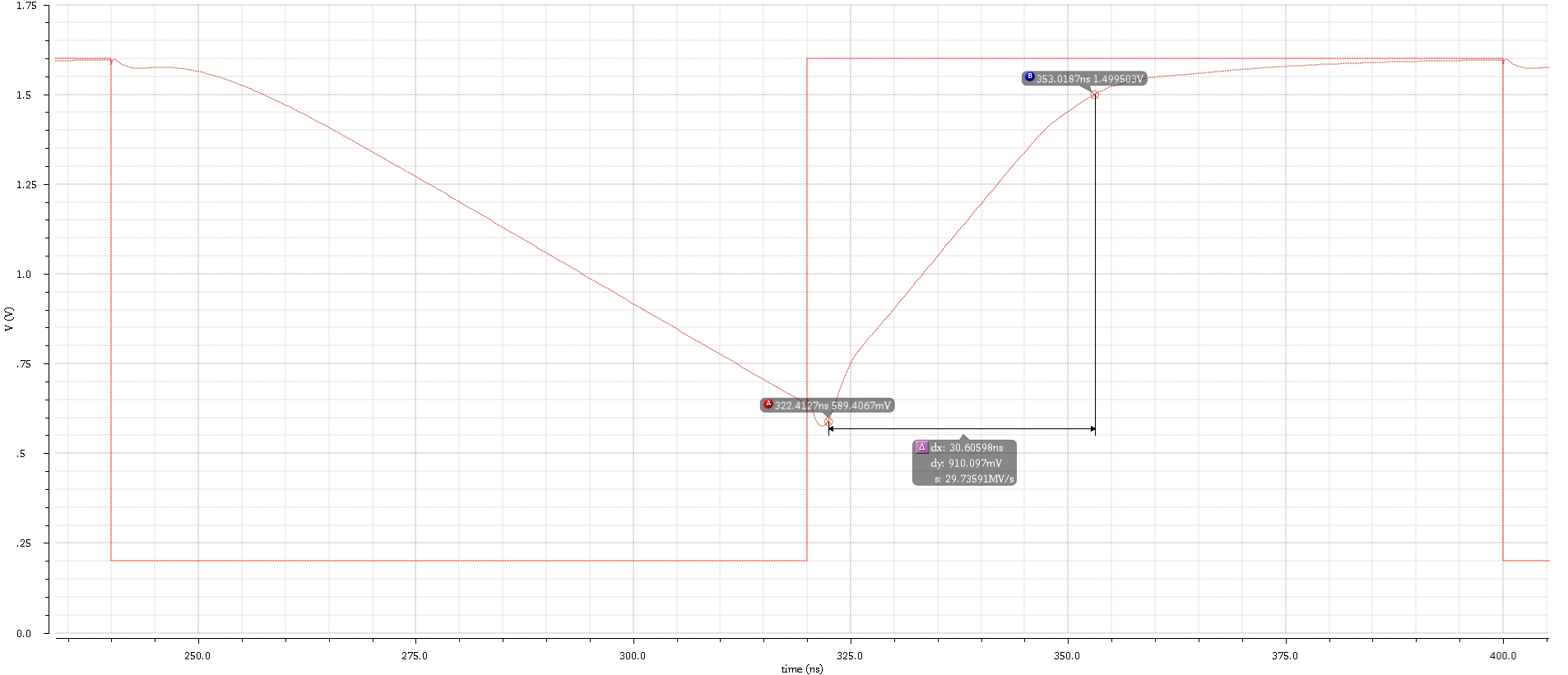}}
\caption{Simulation result of raise slew rate}
\label{slewrateraise}
\end{figure}

\begin{figure}[H]
\centerline{\includegraphics[width=0.5\textwidth]{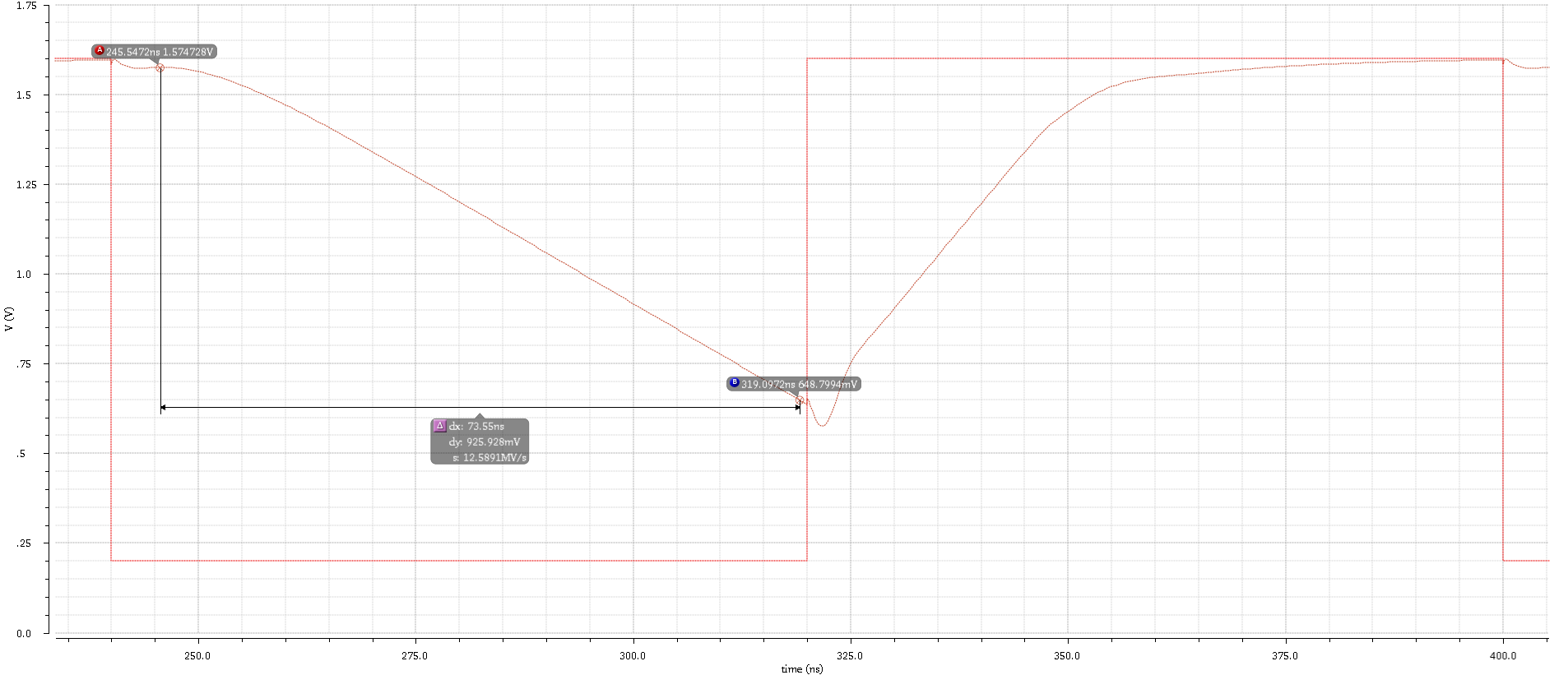}}
\caption{The output voltage swing and fall slew rate}
\label{slewratefall}
\end{figure}

\begin{figure}[H]
\centerline{\includegraphics[width=0.5\textwidth]{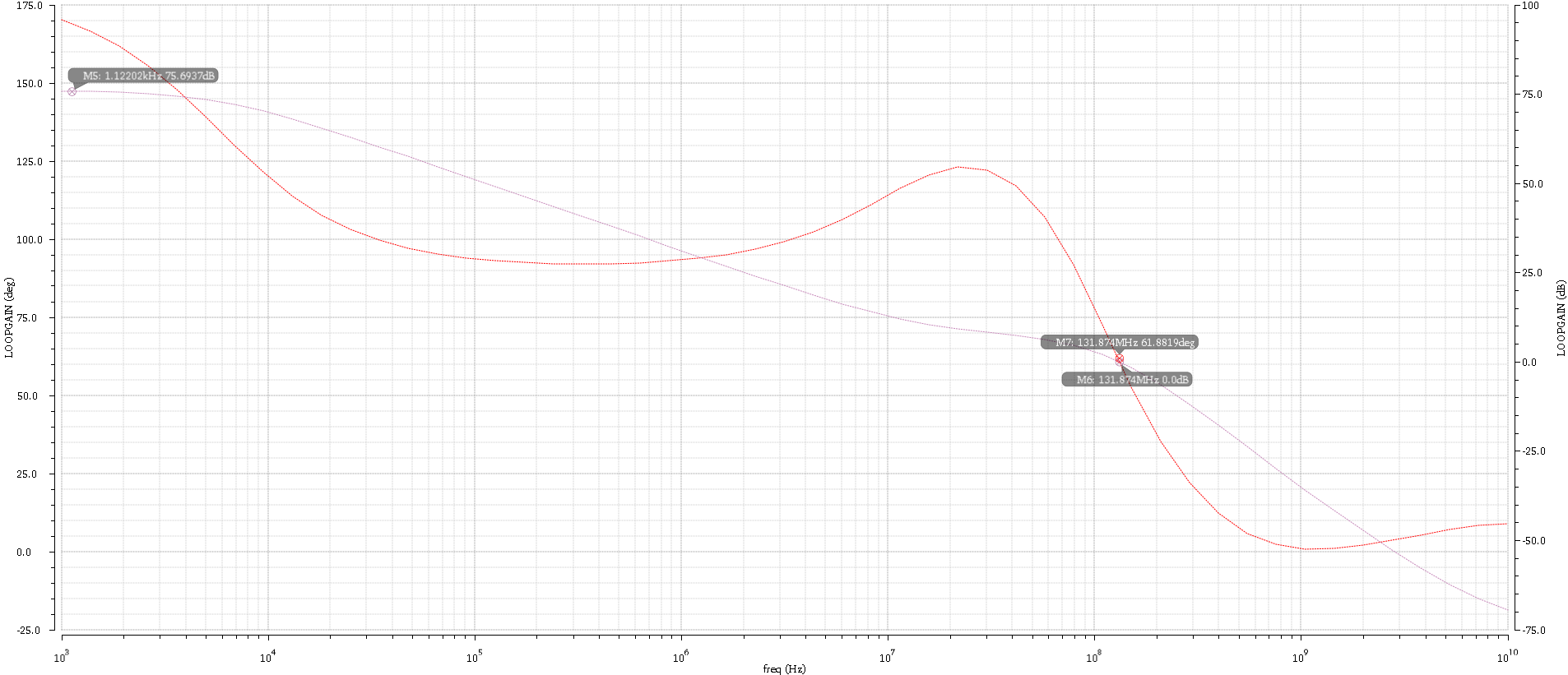}}
\caption{Phase margin and unity gain frequency}
\label{phasemargin}
\end{figure}

% \subfile{tutorial.tex}

%\section{Reference}


\begin{thebibliography}{00}
\bibitem{}T. H. Kim, “6.012 DP: CMOS Integrated Diﬀerential Ampliﬁer,” p. 9.

\bibitem{}R. Sotner, J. Jerabek, R. Prokop, V. Kledrowetz, and J. Polak, “A CMOS Multiplied Input Differential Difference Amplifier: A New Active Device and Its Applications,” Applied Sciences, vol. 7, no. 1, p. 106, Jan. 2017, doi: 10.3390/app7010106.

\bibitem{}A. R. A. El-mon’m and A. W. Abdallah, “CMOS Two-Stage Amplifier Design Approach,” p. 5.

\bibitem{}M. H. Hamzah, A. B. Jambek, and U. Hashim, “Design and analysis of a two-stage CMOS op-amp using Silterra’s 0.13 um technology,” in 2014 IEEE Symposium on Computer Applications and Industrial Electronics (ISCAIE), Penang, Malaysia, Apr. 2014, pp. 55–59, doi: 10.1109/ISCAIE.2014.7010209.

\bibitem{}Shu-Chuan Huang and M. Ismail, “Design of a CMOS differential difference amplifier and its applications in A/D and D/A converters,” in Proceedings of APCCAS’94 - 1994 Asia Pacific Conference on Circuits and Systems, Dec. 1994, pp. 478–483, doi: 10.1109/APCCAS.1994.514597.

\bibitem{}S. Bandyopadhyay, D. Mukherjee, and R. Chatterjee, “Design Of Two Stage CMOS Operational Amplifier in 180nm Technology With Low Power and High CMRR,” p. 9.

\bibitem{}H. Ma, G. H. Nam-Goong, S. Kim, S.-I. Lim, and F. Bien, “Differential Difference Amplifier based Parametric Measurement Unit with Digital Calibration,” JSTS, vol. 18, no. 4, pp. 438–444, Aug. 2018, doi: 10.5573/JSTS.2018.18.4.438.
\end{thebibliography}
\end{document}